\documentclass[conference,10pt]{IEEEtran}
\usepackage{algorithm}
\usepackage[noend]{algpseudocode}

\usepackage{lscape}

\usepackage[english]{babel}
\usepackage[T1]{fontenc}
\usepackage{cite,url,color} 
\usepackage{graphics,amsfonts}
\usepackage{epstopdf}
\usepackage{caption}
\usepackage[pdftex]{graphicx}
\usepackage[cmex10]{amsmath}
\usepackage[utf8]{inputenc}
\usepackage{enumitem}

\usepackage[font=footnotesize]{caption}
\usepackage[font=footnotesize]{subcaption}

\usepackage{tikz}
\usetikzlibrary{automata,positioning,chains,shapes,arrows,patterns,spy}
\usepackage{pgfplots}
\usetikzlibrary{plotmarks}
\newlength\fheight
\newlength\fwidth
\pgfplotsset{compat=newest}
\pgfplotsset{plot coordinates/math parser=false}

\usepackage{array}

\usepackage[top=1.5cm, bottom=2cm, right=1.6cm,left=1.6cm]{geometry}
\usepackage{indentfirst}

\usepackage{times}
\usepackage{glossaries}

\usepackage{breqn}
\usepackage{booktabs}
\usepackage{etoolbox}
\usepackage{placeins}

\newacronym{3gpp}{3GPP}{3rd Generation Partnership Project}
\newacronym{5g}{5G}{5th generation}
\newacronym{aimd}{AIMD}{Additive Increase Multiplicative Decrease}
\newacronym{am}{AM}{Acknowledged Mode}
\newacronym{amc}{AMC}{Adaptive Modulation and Coding}
\newacronym{aqm}{AQM}{Active Queue Management}
\newacronym{awgn}{AGWN}{Additive White Gaussian Noise}
\newacronym{balia}{BALIA}{Balanced Link Adaptation}
\newacronym{bdp}{BDP}{Bandwidth-Delay Product}
\newacronym{bf}{BF}{Beamforming}
\newacronym{cc}{CC}{Congestion Control}
\newacronym{cdf}{CDF}{Cumulative Distribution Function}
\newacronym{cn}{CN}{Core Network}
\newacronym{cqi}{CQI}{Channel Quality Information}
\newacronym{cp}{CP}{Control Plane}
\newacronym{csirs}{CSI-RS}{Channel State Information - Reference Signal}
\newacronym{dc}{DC}{Dual Connectivity}
\newacronym{dce}{DCE}{Direct Code Execution}
\newacronym{dci}{DCI}{Downlink Control Information}
\newacronym{dl}{DL}{Downlink}
\newacronym{dmr}{DMR}{Deadline Miss Ratio}
\newacronym{dmrs}{DMRS}{DeModulation Reference Signal}
\newacronym{e2e}{E2E}{End-to-End}
\newacronym{ecn}{ECN}{Explicit Congestion Notification}
\newacronym{edf}{EDF}{Earliest Deadline First}
\newacronym{enb}{eNB}{evolved Node Base}
\newacronym{epc}{EPC}{Evolved Packet Core}
\newacronym{es}{ES}{Edge Server}
\newacronym{fdma}{FDMA}{Frequency Division Multiple Access}
\newacronym{fdd}{FDD}{Frequency Division Duplexing}
\newacronym[firstplural=Radio Access Technologies (RATs)]{rat}{RAT}{Radio Access Technology}
\newacronym{fs}{FS}{Fast Switching}
\newacronym{ftp}{FTP}{File Transfer Protocol}
\newacronym{gnb}{gNB}{Next Generation Node Base}
\newacronym{harq}{HARQ}{Hybrid Automatic Repeat reQuest}
\newacronym{hetnet}{HetNet}{Heterogeneous Network}
\newacronym{hh}{HH}{Hard Handover}
\newacronym{hol}{HOL}{Head-of-Line}
\newacronym{ia}{IA}{Initial Access}
\newacronym{iot}{IoT}{Internet of Things}
\newacronym{los}{LOS}{Line of Sight}
\newacronym{lte}{LTE}{Long Term Evolution}
\newacronym{m2m}{M2M}{Machine to Machine}
\newacronym{mac}{MAC}{Medium Access Control}
\newacronym{mc}{MC}{Multi-Connectivity}
\newacronym{mcs}{MCS}{Modulation and Coding Scheme}
\newacronym{mec}{MEC}{Mobile Edge Cloud}
\newacronym{mi}{MI}{Mutual Information}
\newacronym{mimo}{MIMO}{Multiple Input, Multiple Output}
\newacronym{mmwave}{mmWave}{millimeter wave}
\newacronym{mptcp}{MPTCP}{Multipath TCP}
\newacronym{mr}{MR}{Maximum Rate}
\newacronym{mss}{MSS}{Maximum Segment Size}
\newacronym{mtd}{MTD}{Machine-Type Device}
\newacronym{mtu}{MTU}{Maximum Transmission Unit}
\newacronym{nfv}{NFV}{Network Function Virtualization}
\newacronym{nlos}{NLOS}{Non Line of Sight}
\newacronym{nr}{NR}{New Radio}
\newacronym{ofdm}{OFDM}{Orthogonal Frequency Division Multiplexing}
\newacronym{pdcch}{PDCCH}{Physical Downlonk Control Channel}
\newacronym{pdcp}{PDCP}{Packet Data Convergence Protocol}
\newacronym{pdsch}{PDSCH}{Physical Downlink Shared Channel}
\newacronym{pdu}{PDU}{Packet Data Unit}
\newacronym{pf}{PF}{Proportional Fair}
\newacronym{pgw}{PGW}{Packet Gateway}
\newacronym{phy}{PHY}{Physical}
\newacronym{pbch}{PBCH}{Physical Broadcast Channel}
\newacronym[plural=\gls{mme}s,firstplural=Mobility Management Entities (MMEs)]{mme}{MME}{Mobility Management Entity}
\newacronym{prb}{PRB}{Physical Resource Block}
\newacronym{pss}{PSS}{Primary Synchronization Signal}
\newacronym{pucch}{PUCCH}{Physical Uplink Control Channel}
\newacronym{pusch}{PUSCH}{Physical Uplink Shared Channel}
\newacronym{rach}{RACH}{Random Access Channel}
\newacronym{ran}{RAN}{Radio Access Network}
\newacronym{red}{RED}{Random Early Detection}
\newacronym{rlc}{RLC}{Radio Link Control}
\newacronym{rrc}{RRC}{Radio Resource Control}
\newacronym{rrm}{RRM}{Radio Resource Management}
\newacronym{rr}{RR}{Round Robin}
\newacronym{rs}{RS}{Remote Server}
\newacronym{rsrp}{RSRP}{Reference Signal Received Power}
\newacronym{rss}{RSS}{Received Signal Strength}
\newacronym{rtt}{RTT}{Round Trip Time}
\newacronym{rw}{RW}{Receive Window}
\newacronym{rx}{RX}{Receiver}
\newacronym{sa}{SA}{standalone}
\newacronym{sack}{SACK}{Selective Acknowledgment}
\newacronym{sap}{SAP}{Service Access Point}
\newacronym{sch}{SCH}{Secondary Cell Handover}
\newacronym{scoot}{SCOOT}{Split Cycle Offset Optimization Technique}
\newacronym{sdma}{SDMA}{Spatial Division Multiple Access}
\newacronym{sinr}{SINR}{Signal to Interference plus Noise Ratio}
\newacronym{sm}{SM}{Saturation Mode}
\newacronym{snr}{SNR}{Signal to Noise Ratio}
\newacronym{son}{SON}{Self-Organizing Network}
\newacronym{ss}{SS}{Synchronization Signal}
\newacronym{srs}{SRS}{Sounding Reference Signal}
\newacronym{sss}{SSS}{Secondary Synchronization Signal}
\newacronym{tb}{TB}{Transport Block}
\newacronym{tcp}{TCP}{Transmission Control Protocol}
\newacronym{tdd}{TDD}{Time Division Duplexing}
\newacronym{tdma}{TDMA}{Time Division Multiple Access}
\newacronym{tfl}{TfL}{Transport for London}
\newacronym{tm}{TM}{Transparent Mode}
\newacronym{trp}{TRP}{Transmitter Receiver Pair}
\newacronym{tti}{TTI}{Transmission Time Interval}
\newacronym{ttt}{TTT}{Time-to-Trigger}
\newacronym{tx}{TX}{Transmitter}
\newacronym{ue}{UE}{User Equipment}
\newacronym{ul}{UL}{Uplink}
\newacronym{uml}{UML}{Unified Modeling Language}
\newacronym{um}{UM}{Unacknowledged Mode}
\newacronym{utc}{UTC}{Urban Traffic Control}
\newacronym{vm}{VM}{Virtual Machine}
\newacronym{rsrq}{RSRQ}{Reference Signal Received Quality}
\newacronym{rssi}{RSSI}{Received Signal Strength Indicator}
\newacronym{crs}{CRS}{Cell Reference Signal}
\newacronym{vss}{VSS}{Video Streaming Server}
\newacronym{rlnc}{RLNC}{Random Linear Network Coding}
\newacronym{ge}{GE}{Gaussian Elimination}
\newacronym{now}{NOW}{Non Overlapping Window}
\newacronym{svc}{SVC}{Scalable Video Coding}
\newacronym{avc}{AVC}{Advanced Video Coding}
\newacronym{gop}{GOP}{Group of Pictures}
\newacronym{nalu}{NALU}{Network Abstraction Layer Unit}
\newacronym{psnr}{PSNR}{Peak Signal to Noise Ratio}
\newacronym{mse}{MSE}{Mean Square Error}

\IEEEoverridecommandlockouts

\begin{document}

\tikzstyle{startstop} = [rectangle, rounded corners, minimum width=2cm, minimum height=0.5cm,text centered, draw=black]
\tikzstyle{io} = [trapezium, trapezium left angle=70, trapezium right angle=110, minimum width=3cm, minimum height=1cm, text centered, draw=black]
\tikzstyle{process} = [rectangle, minimum width=2cm, minimum height=0.5cm, text centered, draw=black, alignb=center]
\tikzstyle{decision} = [ellipse, minimum width=2cm, minimum height=1cm, text centered, draw=black]
\tikzstyle{arrow} = [thick,<->,>=stealth]
\tikzstyle{line} = [thick,>=stealth]
\tikzstyle{darrow} = [thick,<->,>=stealth,dashed]
\tikzstyle{sarrow} = [thick,->,>=stealth]
\tikzstyle{larrow} = [line width=0.1mm,->,>=stealth]

\title{Reliable Video Streaming over mmWave\\with Multi Connectivity and Network Coding}

\author{\IEEEauthorblockN{Matteo Drago, Tommy Azzino, Michele Polese, \v{C}edomir Stefanovi\'c$^*$, Michele Zorzi}
\IEEEauthorblockA{Department of Information Engineering, University of Padova, Padova, Italy\\e-mail: \{dragomat, azzinoto, polesemi, zorzi\}@dei.unipd.it
\\
$^*$Department of Electronic Systems, Aalborg University, Denmark - e-mail: cs@cmi.aau.dk
}
\thanks{This work was partially supported by the U.S. Department of Commerce/NIST (Award No. 70NANB17H166) and by the CloudVeneto initiative. The work of \v{C}edomir Stefanovi\'c was partially supported by the ``Visiting Scientist'' program at the University of Padova. The work of Michele Zorzi was partially supported by NYU Wireless.}}

\makeatletter
\patchcmd{\@maketitle}
  {\addvspace{0.5\baselineskip}\egroup}
  {\addvspace{-1\baselineskip}\egroup}
  {}
  {}
\makeatother

\maketitle

\begin{abstract}
The next generation of multimedia applications will require the telecommunication networks to support a higher bitrate than today, in order to deliver virtual reality and ultra-high quality video content to the users.
Most of the video content will be accessed from mobile devices, prompting the provision of very high data rates by next generation (5G) cellular networks.
A possible enabler in this regard is communication at mmWave frequencies, given the vast amount of available spectrum that can be allocated to mobile users; however, the harsh propagation environment at such high frequencies makes it hard to provide a reliable service.
This paper presents a reliable video streaming architecture for mmWave networks, based on multi connectivity and network coding, and evaluates its performance using a novel combination of the ns-3 mmWave module, real video traces and the network coding library Kodo.
The results show that it is indeed possible to reliably stream video over cellular mmWave links, while the combination of multi connectivity and network coding can support high video quality with low latency.
\end{abstract}

\begin{picture}(0,0)(0,-360)
\put(0,0){
\put(0,0){\footnotesize This paper was invited for presentation at the 2018 IEEE International Conference on} 
\put(0,-10){\footnotesize Computing, Networking and Communications (ICNC), March 2018, Maui, Hawaii, USA.}}
\end{picture}

\section{Introduction}\label{sec:intro}
By 2019, $80$\% of the global Internet consumption will be video content, and $66$\% of it will be accessed from mobile devices~\cite{cisco}.
Moreover, virtual reality applications are becoming more popular, requiring very high data rates, in the order of gigabits per second, and low latency, possibly below 10 ms~\cite{zeqi2017furion}.
Therefore, video and multimedia streaming will be an essential application of the next generation of wireless networks (5G) and of one of its key technological components: mmWave communications.
By using the vast amount of available spectrum at these frequencies, future mobile networks will be able to provide much higher cell data rates to the users, satisfying the increasing demand for high data rates. 

The communication at mmWave frequencies, however, introduces several challenges related to the harsh propagation conditions at such high frequencies~\cite{rangan2014potential}.
The propagation loss is much higher compared to the conventional sub-6 GHz band, thus beamforming techniques and a higher network density are needed to overcome the pathloss.
Moreover, mmWave links are sensitive to blockage from a wide range of materials, such as brick and mortar, but also the human body~\cite{lu2012modeling}.
These issues have an impact on the link capacity, and consequently on the overall end-to-end quality of experience perceived by the final user.
For example, there is a 30 dB difference between the received \gls{sinr} in \gls{los} and \gls{nlos}, with a resulting variation of the data rate offered at the physical layer, as well as possible packet loss.

In order to provide high quality video streaming, besides the high data rate available at mmWave frequencies, there is a need for reliability, low packet loss and a stable data rate.
Moreover, live video streaming also demands low latency.
Motivated by these requirements, we propose in this paper a video streaming architecture that can provide reliable transmission over mmWave links coupled with low latency.
The proposed solution exploits (i) multi-connectivity between the \gls{lte} and mmWave \glspl{ran}, to provide continuous coverage with \gls{lte} and high capacity with mmWave, and (ii) network coding, in order to simplify the management of the transmission on multiple links and provide additional robustness.
We evaluate the performance in terms of packet loss, latency and video quality using a novel framework that combines for the first time the ns-3 mmWave simulator~\cite{mezzavilla2017end} with real video traces and a network coding library~\cite{pedersen2011kodo}.
The results verify that the proposed solution provides a high level of video quality with low delays.

The rest of the paper is organized as follows.
Sec.~\ref{sec:sota} provides an overview of the state of the art.
Sec.~\ref{sec:arch} describes the proposed video streaming architecture, and Sec.~\ref{sec:perf} reports the results of the performance evaluation.
Finally, conclusions and possible extensions are provided in Sec.~\ref{sec:conclusion}.

\begin{figure*}[t]
	\centering
	\includegraphics[width=0.8\textwidth]{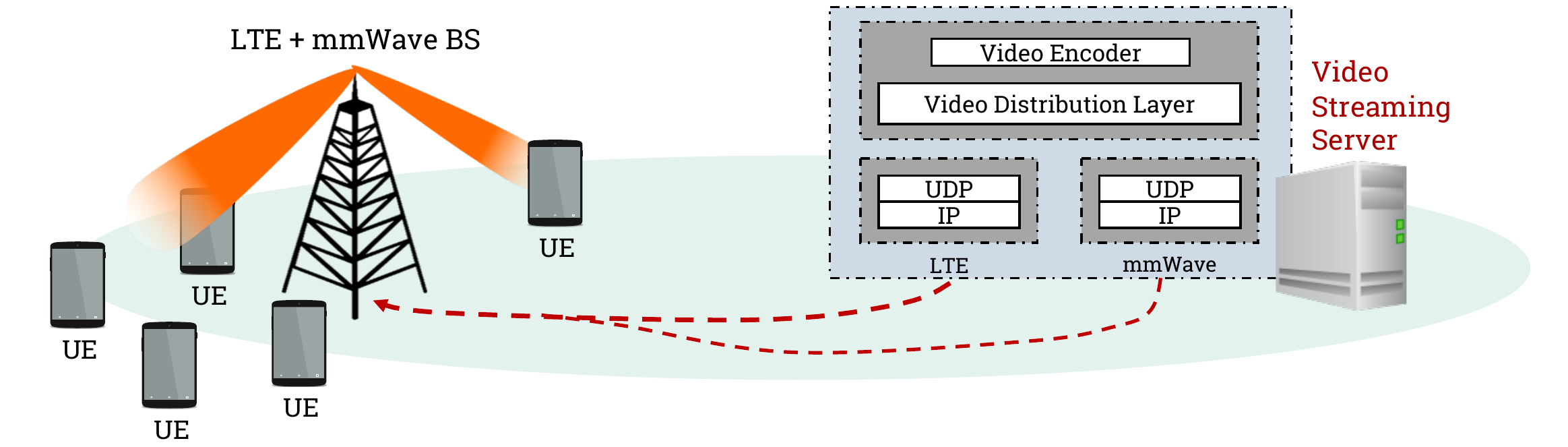}
	\caption{Video Streaming Architecture.}
	\label{fig:videoStreaming}
\end{figure*}

\section{State of the Art}\label{sec:sota}
In the literature, there are both research results and commercial products for indoor applications of video streaming at mmWaves, with a limited range and based on either proprietary technologies~\cite{measy} or IEEE 802.11ad~\cite{nitsche2014ieee}.
In~\cite{singh2008video}, the 60 GHz band is shown as a candidate for the transmission of uncompressed, high quality video up to 3 Gbit/s.
In an indoor environment, mmWave links can also be configured to stream virtual reality content from a local server to the headset~\cite{abari17enabling}.
However, the evaluation of the end-to-end performance of applications in mmWave cellular networks is a research area still in its infancy, given the lack of large mmWave cellular network testbeds or deployments. 

In conventional \gls{lte} cellular networks, network coding has been studied as an enabler of high quality video streaming.
The authors of~\cite{vukobratovic2014random} propose to use it as an error correction technique in the \gls{lte} \gls{ran}, and re-design the \gls{mac} layer in order to use network coding instead of the traditional \gls{harq} mechanism for multimedia traffic.
A similar proposal for WiMAX can be found in~\cite{jin2008is}.
In~\cite{tassi2015resource}, network coding was shown to increase the efficiency of resource allocation when used for video broadcasting in \gls{lte}.

To the best of our knowledge, the combination of multi connectivity and network coding at mmWaves for video streaming has not been studied yet.
Multi connectivity was studied in~\cite{song2012performance} to satisfy the quality of service constraints of video streaming, but without network coding and at sub-6 GHz frequencies. 
On the other hand, a packet-level encoding technique similar to network coding (i.e., the Luby Transform codes) was used on top of UDP on a mmWave link in~\cite{mmMAGIC}, to increase the connection goodput with respect to TCP, but not for video streaming and without multi connectivity.

\section{Architecture}\label{sec:arch}

In this section, we present the proposed architecture for video streaming in a mmWave cellular network scenario.
As shown in Fig.~\ref{fig:videoStreaming}, the intelligent distribution policy is located in the \gls{vss}, which can be deployed either in the operator's core network as a caching server, or in the public internet.
A middle layer (called video distribution layer) to manage network coding, any retransmissions, and the multiple interfaces to the different \glspl{ran} is placed between the encoding layer, which generates video frames, and the transport layer. Both UDP and TCP have been used as transport protocols for video distribution: for example the Dynamic Adaptive Streaming over HTTP (DASH) protocol~\cite{stockhammer2011dash} relies on TCP, while the Real Time Streaming Protocol (RSTP)~\cite{RFC7826} can operate on both. In our framework we consider UDP for two reasons.
The first is that the reliability typically offered by TCP is provided in our architecture by network coding at the middle layer, and the second is that the performance of TCP on mmWave links has a number of limitations, mainly related to the bufferbloat phenomenon in \gls{los} to \gls{nlos} transitions (i.e., the latency and the jitter on the link increase) 
and low efficiency in terms of throughout with respect to the data rate achievable by a mmWave link~\cite{zhang2016transport,polese2017tcp}.

In the next paragraphs, we describe how each of the three components of our solution (i.e., multi connectivity, network coding and the video transmission policy) are engineered to yield the best performance for the final user.

\subsection{Multi Connectivity}

The possibility of using multiple network interfaces at the same time is an emerging paradigm in wireless and data center communications~\cite{raiciu2011improving}.
In particular, the modern smartphones are generally equipped with multiple radios and network interfaces.
In this paper, we use multi connectivity at the application layer to communicate using different \glspl{rat}, such as \gls{lte} at sub-6 GHz frequencies and \gls{nr} at mmWave frequencies, in order to benefit from (i) a more reliable end-to-end packet transmission on LTE when the mmWave link is not available, and (ii) the very high data rate of the mmWave connection when the link quality is high enough. 
Moreover, the LTE connection is also used to send feedback messages from the \gls{ue} to the \gls{vss} to signal the availability and the quality of the mmWave link.
Contrary to what is proposed in~\cite{polese2017improved,dasilva}, the integration is not in the cellular protocol stack but at a higher layer, similarly to what happens with Multipath TCP~\cite{rfc6824}. This makes the deployments of multi connectivity independent of the choices of the network operator, and, as long as a final user can use multiple independent radios in its smartphone, an over-the-top video content provider can exploit multi connectivity in its streaming application.

\subsection{Network Coding}

Network coding is a packet-level encoding technique that combines source packets using algebraic operations in order to increase the resilience with respect to packet loss in an efficient way~\cite{ahlswede2000network}.
In our architecture, we rely on the rateless version of \gls{rlnc}~\cite{ho2006random}, as it provides a good trade-off between bandwidth efficiency, complexity and delay, compared to other network coding or forward error correction strategies~\cite{NC-areview}.
The network coding library chosen for this paper is Kodo~\cite{pedersen2011kodo}.

With \gls{rlnc}, the packets generated by the video encoding layer are grouped into generations, i.e., sets of $K$ packets meant to be encoded together, where $K$ is the generation size.
For each generation, coded packets are obtained as independent random linear combinations of the $K$ packets, where symbols, coefficients and all operations are defined in the Galois field with $q$ elements, $\mathbb{F}_q$. As a result, every encoded packet is an equally useful representation of the packets from the generation, such that the decoder is able to decode the original information using any combination of (slightly more than) $K$ encoded packets. 
The number of encoded packets that can be generated from $K$ packets, i.e., the \gls{rlnc} code rate is not fixed.
If some packets are lost on the mmWave link, it is possible to produce newly encoded packets without re-encoding and retransmitting the whole generation.
This is the rateless property of the encoding scheme.

When an encoded packet is produced, it can be immediately transmitted.
The decoder collects encoded packets, and needs to receive at least $K$ packets to attempt a successful decoding.
At the decoder side, the original packets are retrieved through Gaussian elimination, by constructing a decoding matrix with the linearly independent encoded packets that have been successfully received.
Since the encoding coefficients are randomly chosen, it is not guaranteed that each encoded packet will be linearly independent of the others, and thus that the original payloads will be re-constructed given $K$ encoded packets.
In order to increase the decoding probability, in our design we send $N\ge K$ encoded packets and start decoding on-the-fly as soon as $K$ packets are received. 

There are trade-offs between (i) the latency and the decoding probability, which both increase with the generation size $K$, and (ii) the decoding complexity and the decoding probability, which both increase with the field size $q$~\cite{heide2011code}.
We test two different configurations: configuration $LC$ ($K=40$, $q=4$), which offers low latency and decoding overhead at the cost of a lower probability of successful decoding; and configuration $HC$ ($K=100$, $q=8$) where the latency, overhead, and also the probability of successful decoding are increased. 


Finally, network coding simplifies the management of multi connectivity, since the retransmissions do not need to be performed on the path in which the lost packet was originally transmitted, but the best available one can be used when needed.
In order to protect from both unsuccessful decoding and packet loss on the wireless link, $N$ is set to respectively $1.2K$ or $1.1K$ when the mmWave or the LTE link is used.
Transmissions of additional, newly encoded packets can be triggered, up to a maximum number of 5 attempts.\footnote{Strictly speaking, the proposed scheme is not fully rateless, as the number of generated encoded packets has an upper limit due to latency constraints.}

\subsection{Video Encoding Policy}\label{sec:video}

The video is encoded using the H.264/\gls{avc} standard~\cite{avc} with the \gls{svc} extension~\cite{SVC}.
This framework provides the possibility of avoiding the transmission of some portions of the video bit stream in order to adapt the source rate to the channel capacity or to the needs of the end users: this property has been referred to as \textit{scalability}.
The 
source content can be divided into subsets with a reduced picture size (spatial scalability) or lower frame rate (temporal scalability).
In the time domain, it is possible to identify key frames that will carry most of the content, and enhancement frames that are placed between two key frames and can be discarded (with a loss of quality). The different kinds of frames belong to different temporal layers.
The key frames are part of the temporal base layer, and two of these frames together with a set of enhancement frames form a \gls{gop}.
In the spatial domain, the scalability makes it possible to code two or more versions of the same video at different resolutions in a unique bit stream, which is therefore composed of different layers corresponding to different spatial resolutions (i.e., a spatial layer).
According to the H.264/\gls{avc} standard, the bit stream generated by the encoder is divided into \glspl{nalu}, each with a payload containing a portion of the encoded video frame.
Each \gls{nalu} is then split into packets of size $P=1000$~bytes, which are forwarded to the network coding layer.
According to the \gls{now} policy~\cite{UEP-RLC}, the network coding layer maps packets of different \glspl{nalu} into different generations, so that the encoding is independent for each \gls{nalu}.

In this paper, we consider a 50 Hz video with \gls{gop} of 16 frames~\cite{SVC}, 5 temporal layers, and 2 spatial layers at a resolution of 720p (base layer) and 1080p (enhancement layer).


\section{Performance Evaluation}\label{sec:perf}

\subsection{Simulation Setup}
The performance evaluation is carried out using the mmWave module~\cite{wns3_16,mezzavilla2017end} of the ns-3 simulator, which was integrated with the Kodo network coding library~\cite{pedersen2011kodo} and several tools to process the video traces. 

The ns-3 mmWave module is equipped with a full 3GPP-like cellular stack for both LTE and mmWave.
It enables the simulation of an end-to-end network, from the \acrlong{vss} to the \gls{ue}, with a realistic channel model that is based on the 3GPP specifications~\cite{zhang2017ns} and the possibility of adding obstacles to model \gls{nlos} links.
The simulated protocol stack includes \gls{phy} and \gls{mac} layers based on a low-latency \gls{tdd} design, with optional link-layer retransmissions at the \gls{mac} layer (using \gls{harq}) and at the \gls{rlc} layer.
The scenario contains a single cell with radius equal to 100 m and 5 users, 2 in \gls{los} and 3 in \gls{nlos}, which move at a random speed between 2 and 4 m/s around fixed positions in the cell.
The main parameters of the simulations are given in Table~\ref{table:simparams}. 

\begin{table}[b]
	\begin{center}
		\begin{tabular}{@{}ll@{}}
			\toprule
			Parameters & Value \\
			\midrule
			LTE carrier frequency (DL) & 2.1 GHz \\
			LTE carrier frequency (UL) & 1.9 GHz \\
			LTE bandwitdh & 20 MHz \\
			LTE downlink TX power $P_{TX}$ & 43 dBm \\
			mmWave carrier frequency & 28 GHz \\
			mmWave bandwidth & 1 GHz \\
			mmWave $P_{TX}$ & 30 dBm \\
    		3GPP Channel Scenario & Urban Micro \\
			mmWave SNR Outage Threshold & -5 dB \\
		    RLC buffer size $B_{RLC}$ & 20 MB \\
    		RLC reordering timer & 1 ms \\ 
    		RLC Buffer Status Report timer & 2 ms \\ 
    		Number of UEs & 5 (2 \gls{los}, 3 \gls{nlos})\\
    		\gls{vss}-\gls{ue} latency & 10 ms \\
			\bottomrule
		\end{tabular}
	\end{center}
	\caption{Main simulation parameters}
	\label{table:simparams}
\end{table}

In order to provide a realistic video streaming model, the chosen video sample is first encoded in the format specified in Sec.~\ref{sec:video}, from which a bit stream is then generated using the JSVM software~\cite{JSVM}.
Using our extension of the tool provided in~\cite{SVEF}, the bit stream is adapted to the processing in ns-3. The \glspl{nalu} are then handled by the video distribution layer and transmitted in the simulation, and, at the \gls{ue}, the correctly received frames are first buffered and then played-out. The play-out action in the simulation corresponds to writing the frame-related information in an output trace, which is then processed with the tool in~\cite{SVEF} in order to allow the video reconstruction and quality evaluation with FFmpeg~\cite{ffmpeg}. The video buffer considered in the simulation has a memory of 25 frames, i.e., 500 ms of video.

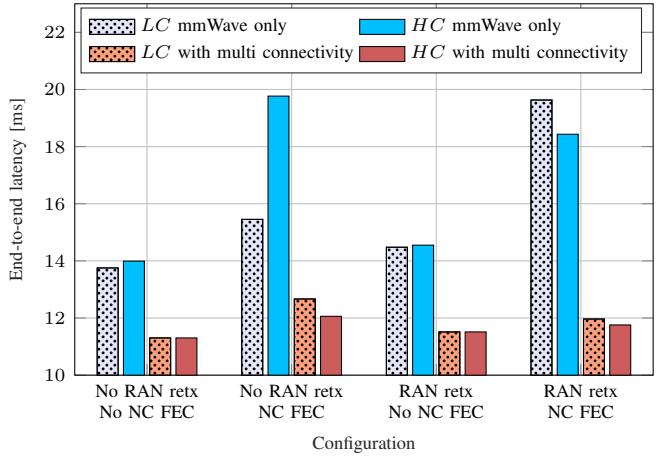
\begin{figure}[t]
	\centering
	\setlength\fwidth{0.9\columnwidth}
	\setlength\fheight{0.55\columnwidth}
%
%
\definecolor{mycolor1}{rgb}{0.20810,0.16630,0.52920}%
\definecolor{mycolor2}{rgb}{0.02650,0.61370,0.81350}%
\definecolor{mycolor3}{rgb}{0.64730,0.74560,0.41880}%
\definecolor{mycolor4}{rgb}{0.97630,0.98310,0.05380}%
\definecolor{lavender}{rgb}{0.9020,0.9020,0.9804}%
\definecolor{lightskyblue}{rgb}{0.6784,0.8471,0.9020}%
\definecolor{deepskyblue}{rgb}{0,0.7490,1}%
\definecolor{steelblue}{rgb}{0.2745,0.5098,0.7059}%
\definecolor{blue}{rgb}{0,0,1}%
\definecolor{royalblue}{rgb}{0.2549,0.4118,0.8824}%

\definecolor{gainsboro}{rgb}{0.8627,0.8627,0.8627}%
\definecolor{darkslategrey}{rgb}{0.1843,0.3098,0.3098}%
\definecolor{gray}{rgb}{0.5,0.5,0.5}%

\definecolor{lightcoral}{rgb}{0.9412,0.5020,0.5020}%
\definecolor{indianred}{rgb}{0.8039,0.3608,0.3608}%
\definecolor{lightsalmon}{rgb}{1.0000,0.6275,0.4784}%
\definecolor{darksalmon}{rgb}{0.9137,0.5882,0.4784}%
\begin{tikzpicture}
\pgfplotsset{every tick label/.append style={font=\scriptsize}}

\begin{axis}[%
width=0.951\fwidth,
height=\fheight,
at={(0\fwidth,0\fheight)},
scale only axis,
bar shift auto,
xmin=0.5,
xmax=4.5,
xtick={1,2,3,4},
xticklabel style={align=center},
xticklabels={{No \gls{ran} retx\\No NC FEC},{No \gls{ran} retx\\NC FEC},{\gls{ran} retx\\No NC FEC},{\gls{ran} retx\\NC FEC}},
ylabel style={font=\scriptsize\color{white!15!black}},
ylabel={End-to-end latency [ms]},
xlabel style={font=\scriptsize\color{white!15!black}},
xlabel={Configuration},
ymin=10,
ymax=23,
axis background/.style={fill=white},
xmajorgrids,
ymajorgrids,
legend columns=2,
legend style={at={(0.01, 0.99)},anchor=north west, font=\scriptsize, legend cell align=left, align=left, draw=white!15!black}
]
\addplot[ybar, postaction={pattern=crosshatch dots}, bar width=0.145, fill=lavender, draw=black, area legend] table[row sep=crcr] {%
1	13.7532627080305\\
3	14.4773036313388\\
2	15.4526183711381\\
4	19.6311499685597\\
};
\addplot[forget plot, color=white!15!black] table[row sep=crcr] {%
0.5	0\\
4.5	0\\
};
\addlegendentry{$LC$ mmWave only}

\addplot[ybar, bar width=0.145, fill=deepskyblue, draw=black, area legend] table[row sep=crcr] {%
1	13.9919837472568\\
3	14.5504021010274\\
2	19.7675133327236\\
4	18.4332122430885\\
};
\addplot[forget plot, color=white!15!black] table[row sep=crcr] {%
0.5	0\\
4.5	0\\
};
\addlegendentry{$HC$ mmWave only}

\addplot[ybar, bar width=0.145, postaction={pattern=crosshatch dots}, fill=lightsalmon, draw=black, area legend] table[row sep=crcr] {%
1	11.3032302403071\\
3	11.5136525691795\\
2	12.6670261025164\\
4	11.9646700959973\\
};
\addplot[forget plot, color=white!15!black] table[row sep=crcr] {%
0.5	0\\
4.5	0\\
};
\addlegendentry{$LC$ with multi connectivity}

\addplot[ybar, bar width=0.145, fill=indianred, draw=black, area legend] table[row sep=crcr] {%
1	11.3032302403071\\
3	11.5136525691795\\
2	12.0604838240262\\
4	11.7575190058516\\
};
\addplot[forget plot, color=white!15!black] table[row sep=crcr] {%
0.5	0\\
4.5	0\\
};
\addlegendentry{$HC$ with multi connectivity}

\end{axis}
\end{tikzpicture}%
	\caption{Latency for different configurations of the packet transmission policy. NC FEC/\gls{ran} retx means that network coding error correction/\gls{rlc} and \gls{harq} retransmission are used, no NC FEC/no \gls{ran} retx otherwise. The y-axis starts from 10 ms, i.e., the value of the \gls{vss}-\gls{ue} latency.}
	\label{fig:latency}
\end{figure}

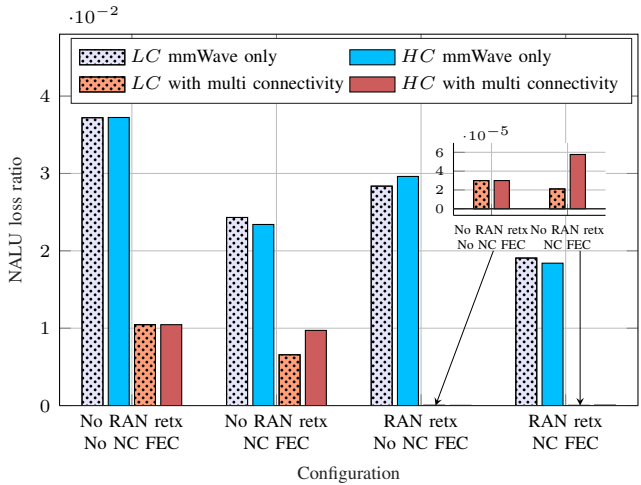
\begin{figure}[t]
	\centering
	\setlength\fwidth{0.9\columnwidth}
	\setlength\fheight{0.55\columnwidth}
%
%
\definecolor{mycolor1}{rgb}{0.20810,0.16630,0.52920}%
\definecolor{mycolor2}{rgb}{0.02650,0.61370,0.81350}%
\definecolor{mycolor3}{rgb}{0.64730,0.74560,0.41880}%
\definecolor{mycolor4}{rgb}{0.97630,0.98310,0.05380}%
\definecolor{lavender}{rgb}{0.9020,0.9020,0.9804}%
\definecolor{lightskyblue}{rgb}{0.6784,0.8471,0.9020}%
\definecolor{deepskyblue}{rgb}{0,0.7490,1}%
\definecolor{steelblue}{rgb}{0.2745,0.5098,0.7059}%
\definecolor{blue}{rgb}{0,0,1}%
\definecolor{royalblue}{rgb}{0.2549,0.4118,0.8824}%

\definecolor{gainsboro}{rgb}{0.8627,0.8627,0.8627}%
\definecolor{darkslategrey}{rgb}{0.1843,0.3098,0.3098}%
\definecolor{gray}{rgb}{0.5,0.5,0.5}%

\definecolor{lightcoral}{rgb}{0.9412,0.5020,0.5020}%
\definecolor{indianred}{rgb}{0.8039,0.3608,0.3608}%
\definecolor{lightsalmon}{rgb}{1.0000,0.6275,0.4784}%
\definecolor{darksalmon}{rgb}{0.9137,0.5882,0.4784}%
\begin{tikzpicture}
\pgfplotsset{every tick label/.append style={font=\scriptsize}}

\begin{axis}[%
width=0.951\fwidth,
height=\fheight,
at={(0\fwidth,0\fheight)},
scale only axis,
bar shift auto,
xmin=0.5,
xmax=4.5,
xtick={1,2,3,4},
xticklabel style={align=center},
xticklabels={{No \gls{ran} retx\\No NC FEC},{No \gls{ran} retx\\NC FEC},{\gls{ran} retx\\No NC FEC},{\gls{ran} retx\\NC FEC}},
ylabel style={font=\scriptsize\color{white!15!black}},
ylabel={NALU loss ratio},
xlabel style={font=\scriptsize\color{white!15!black}},
xlabel={Configuration},
ymin=0,
ymax=0.048,
axis background/.style={fill=white},
xmajorgrids,
ymajorgrids,
legend columns=2,
legend style={at={(0.99, 0.99)},anchor=north east, font=\scriptsize, legend cell align=left, align=left, draw=white!15!black}
]

\addplot[ybar, postaction={pattern=crosshatch dots}, bar width=0.145, fill=lavender, draw=black, area legend] table[row sep=crcr] {%
1	0.0372029281579843\\
3	0.028361139484735\\
2	0.0243105209397344\\
4	0.0190693451367609\\
};
\addplot[forget plot, color=white!15!black] table[row sep=crcr] {%
0.5	0\\
4.5	0\\
};
\addlegendentry{$LC$ mmWave only}

\addplot[ybar, bar width=0.145, fill=deepskyblue, draw=black, area legend] table[row sep=crcr] {%
1	0.0372329033153003\\
3	0.0296152032181995\\
2	0.0234074074074074\\
4	0.0184047718130115\\
};
\addplot[forget plot, color=white!15!black] table[row sep=crcr] {%
0.5	0\\
4.5	0\\
};
\addlegendentry{$HC$ mmWave only}

\addplot[ybar, bar width=0.145, postaction={pattern=crosshatch dots}, fill=lightsalmon, draw=black, area legend] table[row sep=crcr] {%
1	0.0104558191149951\\
3	2.99625468164794e-05\\
2	0.00656845609654598\\
4	2.10847551671522e-05\\
};
\addplot[forget plot, color=white!15!black] table[row sep=crcr] {%
0.5	0\\
4.5	0\\
};
\addlegendentry{$LC$ with multi connectivity}

\addplot[ybar, bar width=0.145, fill=indianred, draw=black, area legend] table[row sep=crcr] {%
1	0.0104558191149951\\
3	2.99625468164794e-05\\
2	0.00971785268414482\\
4	5.7705645720627e-05\\
};
\addplot[forget plot, color=white!15!black] table[row sep=crcr] {%
0.5	0\\
4.5	0\\
};
\addlegendentry{$HC$ with multi connectivity}

\coordinate (pt) at (axis cs:3.75,0.012);
\draw [larrow] (axis cs:3.5,0.02) -- (3.1,0);
\draw [larrow] (axis cs:4.1,0.02) -- (4.1,0);

\end{axis}

\node[pin={[pin edge={lavender,ultra thin}]90:{%
    \begin{tikzpicture}[trim axis left,trim axis right]
    \pgfplotsset{every tick label/.append style={font=\tiny}}
    \begin{axis}[
    scale only axis,
bar shift auto,
width=0.25\fwidth,
height=0.195\fheight,
axis x line*=bottom,
axis y line*=left, 
xmin=2.5,
xmax=4.5,
xtick={3,4},
xticklabel style={align=center},
xticklabels={{No \gls{ran} retx\\No NC FEC},{No \gls{ran} retx\\NC FEC},{\gls{ran} retx\\No NC FEC},{\gls{ran} retx\\NC FEC}},
xmajorgrids,
ymajorgrids,
ymax=7e-5,
axis background/.style={fill=white},
    ]


\addplot[ybar, bar width=0.2, postaction={pattern=crosshatch dots}, fill=lightsalmon, draw=black, area legend] table[row sep=crcr] {%
3	2.99625468164794e-05\\
4	2.10847551671522e-05\\
};
\addplot[forget plot, color=white!15!black] table[row sep=crcr] {%
0.5	0\\
4.5	0\\
};

\addplot[ybar, bar width=0.2, fill=indianred, draw=black, area legend] table[row sep=crcr] {%
3	2.99625468164794e-05\\
4	5.7705645720627e-05\\
};
\addplot[forget plot, color=white!15!black] table[row sep=crcr] {%
0.5	0\\
4.5	0\\
};

    \end{axis}
    \end{tikzpicture}%
}}] at (pt) {};

\end{tikzpicture}%
	\caption{\gls{nalu} loss ratio for different configurations of the packet transmission policy. NC FEC/\gls{ran} retx means that network coding error correction/\gls{rlc} and \gls{harq} retransmission are used, no NC FEC/no \gls{ran} retx otherwise.}
	\label{fig:nalu_linear}
\end{figure}

\begin{figure}[t]
	\centering
	\setlength\fwidth{0.9\columnwidth}
	\setlength\fheight{0.55\columnwidth}
	\definecolor{mygreen}{RGB}{28,172,0} 
\definecolor{mylilas}{RGB}{170,55,241}
\begin{tikzpicture}
\pgfplotsset{every tick label/.append style={font=\scriptsize}}
	
\begin{axis}[
width=0.951\fwidth,
height=\fheight,
at={(0\fwidth,0\fheight)},
scale only axis,
xlabel style={font=\scriptsize\color{white!15!black}},
xlabel={End-to-end latency [ms]},
ylabel style={font=\scriptsize\color{white!15!black}},
ylabel={PSNR [dB]},
xmajorgrids,
ymajorgrids,
xminorgrids,
yminorgrids,
minor tick num=1,
legend style={font=\scriptsize, at={(0.99,0.4)},anchor=south east, legend cell align=left, align=left, draw=white!15!black},
xmin=10,
ymin=0,
]

\addplot[
visualization depends on={\thisrow{nodes}\as\myvalue},
scatter/classes={
	A0={mark=x,black},%
	D0={mark=x,orange!80!black},
	AF0={mark=x,green!80!black},
	DF0={mark=x,blue!80!black},
	A1={mark=asterisk,black},%
	D1={mark=asterisk,orange!80!black},
	AF1={mark=asterisk,green!80!black},
	DF1={mark=asterisk,blue!80!black}
},
scatter, only marks,
scatter src=explicit symbolic,
nodes near coords*={\scriptsize\pgfmathprintnumber[int detect]\myvalue},forget plot]
table[x=x,y=y,meta=label]
{./figures/PSNR-A.dat};

\addplot[
visualization depends on={\thisrow{nodes}\as\myvalue},
scatter/classes={
	A0={mark=text,black},%
	D0={mark=x,orange!80!black},
	AF0={mark=x,green!80!black},
	DF0={mark=x,blue!80!black},
	A1={mark=asterisk,black},%
	D1={mark=asterisk,orange!80!black},
	AF1={mark=asterisk,green!80!black},
	DF1={mark=asterisk,blue!80!black}
},
scatter, only marks,
scatter src=explicit symbolic,
nodes near coords*={\scriptsize\pgfmathprintnumber[int detect]\myvalue},forget plot]
table[x=x,y=y,meta=label]
{./figures/PSNR-D.dat};

\addplot[
visualization depends on={\thisrow{nodes}\as\myvalue},
scatter/classes={
	A0={mark=x,black},%
	D0={mark=x,orange!80!black},
	AF0={mark=x,green!80!black},
	DF0={mark=x,blue!80!black},
	A1={mark=asterisk,black},%
	D1={mark=asterisk,orange!80!black},
	AF1={mark=asterisk,green!80!black},
	DF1={mark=asterisk,blue!80!black}
},
scatter, only marks,
scatter src=explicit symbolic,
point meta=explicit symbolic,
nodes near coords*={\scriptsize\pgfmathprintnumber[int detect]\myvalue},forget plot]
table[x=x,y=y,meta=label]
{./figures/PSNR-AF.dat};

\addplot[
visualization depends on={\thisrow{nodes}\as\myvalue},
scatter/classes={
	A0={mark=x,black},%
	D0={mark=x,orange!80!black},
	AF0={mark=x,green!80!black},
	DF0={mark=x,blue!80!black},
	A1={mark=asterisk,black},%
	D1={mark=asterisk,orange!80!black},
	AF1={mark=asterisk,green!80!black},
	DF1={mark=asterisk,blue!80!black}
},
scatter, only marks,
scatter src=explicit symbolic,
nodes near coords*={\scriptsize\pgfmathprintnumber[int detect]\myvalue},forget plot]
table[x=x,y=y,meta=label]
{./figures/PSNR-DF.dat};

\addplot[scatter, only marks, mark=x]
table[row sep=crcr] {%
0 0\\
};
\addlegendentry{1 - No \gls{ran} retx, no NC FEC}

\addplot[scatter, only marks, mark=asterisk]
table[row sep=crcr] {%
0 0\\
};
\addlegendentry{2 - \gls{ran} retx, NC FEC}





\node[at={(13.7532627080305,26.7508477842004)}, pin={[pin distance=1mm]below:\scriptsize $LC$ mmWave}]{};
\node[at={(13.9919837472568,26.7527306967985)}, pin={[pin distance=1mm]right:\scriptsize $HC$ mmWave}]{};
\node[at={(11.3032302403071,27.7779259259259)}, pin={[pin distance=1mm, align=center]below:\scriptsize $LC$ and $HC$\\\scriptsize with multi}]{};
\node[at={(11.7575190058516,96.7142407407406)}, pin={[pin distance=1mm, align=center]below:\scriptsize $HC$ multi}]{};
\node[at={(11.9646700959973,99.2062592592591)}, pin={[pin distance=1mm, align=center]right:\scriptsize $LC$ multi}]{};
\node[at={(19.6311499685597,94.0708952380951)}, pin={[pin distance=1mm]below:\scriptsize $LC$ mmWave}]{};
\node[at={(18.5186276254163,88.8574859287053)}, pin={[pin distance=1mm]left:\scriptsize $HC$ mmWave}]{};

\end{axis}

\end{tikzpicture}
	\caption{PSNR of the spatial base layer vs. latency for the configuration with no \gls{ran} retransmission and no NC FEC (1) and the one with both \gls{ran} and NC FEC (2).}	
	\label{fig:psnr}
\end{figure}
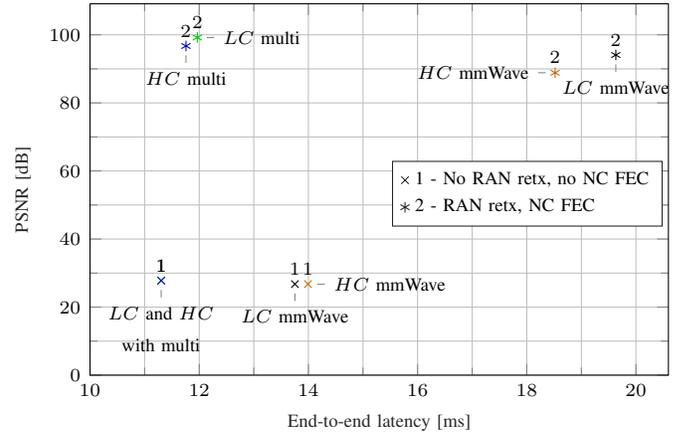

\subsection{Results} 

The metrics considered in this performance evaluation, obtained via Monte Carlo simulations with 90 independent runs, are (i) the \gls{nalu} loss ratio; (ii) the application layer latency, i.e., the delay between the time at which the video frame is generated at the \gls{vss} and when it is consumed by the application at the \gls{ue}; and (iii) the average frame \gls{psnr}.
The \gls{psnr} is a measure of the quality of reconstructed video that is inversely proportional to the \gls{mse} of the received frame $R$ with respect to the original frame $I$.
Given the frame width $W$ and height $H$ in pixels, the \gls{psnr} for frame $n$ is given by~\cite{JSVM}:
\begin{equation}
	PSNR(n) = 10 \log_{10} \frac{WH(2^8-1)^2}{\sum_{w=1}^{W}\sum_{h=1}^{H} [I_n(w, h) - R_n(w,h)]^2}.
\end{equation}
When there are no differences between the reconstructed video frames and the original ones, the \gls{psnr} as in JSVM is assigned the maximum value of $99.99$~dB.

Fig.~\ref{fig:latency} and Fig.~\ref{fig:nalu_linear} compare the end-to-end latency and the \gls{nalu} loss ratio of different configurations.
In particular, two different kinds of error correction to compensate for the packet losses are considered, via link-level retransmissions in the \gls{ran} and/or by transmissions of additional NC packets (henceforth denoted by NC FEC).
When multi connectivity is not used, \gls{ran} retransmissions do not significantly increase the latency, but perform worse than the NC FEC in terms of \gls{nalu} loss reduction with respect to the no error correction case.
The link-level retransmissions are indeed more efficient with respect to single-packet losses in the channel, while NC FEC protects larger chunks of packets and can yield a lower \gls{nalu} loss in case of more extended bursty errors.
The best performance in terms of \gls{nalu} loss when multi connectivity is not used is achieved when combining both \gls{ran} retransmissions and NC FEC, at the price of a modest increase in latency. 

Multi connectivity, however, is the configuration that performs best both for the latency and the \gls{nalu} loss ratio.
In particular, multi connectivity makes it possible to continuously transmit packets even when the mmWave link is in outage, thanks to the \gls{lte} fall back and to the seamless switch enabled by the fact that the \gls{ue} is already connected to both \glspl{ran}.
Therefore, as shown in Fig.~\ref{fig:latency}, 
the average value of the latency when \gls{ran} retransmissions and NC FEC are introduced does not increase significantly, and is in general less than 2 ms higher than the 10 ms delay introduced by the fixed backhaul network.
Additionally, the \gls{nalu} loss with multi connectivity, \gls{ran} retransmissions and NC FEC has very small values (in the order of $10^{-5}$, as shown in Fig.~\ref{fig:nalu_linear}), given that a more reliable \gls{lte} link is used when the mmWave one is in outage. 


Finally, Fig.~\ref{fig:psnr} shows the trade-off between the average end-to-end latency and the average \gls{psnr} of the spatial base layer.\footnote{Note that FFmpeg does not support spatial scalability for video decoding, therefore it is possible to reconstruct only the spatial base layer at 720p. The evaluation of the \gls{psnr} for the combined spatial base and enhancement layers is left for future work.} It can be seen that the \gls{psnr} without error control (points marked with ``1'') is limited to about 26 dB because of the relatively high \gls{nalu} loss ratio, whereas the combination of \gls{ran} retransmissions and network coding error correction (points marked with ``2'') is able to guarantee almost perfect reconstruction (note that 100 dB is the conventional value given by JSVM to error-free packet delivery). The figure also shows that multi connectivity, while not necessarily needed for high \gls{psnr} (which can be achieved even by a stand-alone mmWave network), can be very effective in reducing latency, especially in the presence of error control.

\section{Conclusions}\label{sec:conclusion}
This paper investigates reliable streaming of high-quality video over a mmWave link.
The communication at such high frequencies potentially enables the high data rates needed to support the source rate of high quality videos.
However, challenges are introduced by the mmWave channel instability and the possibility of outages.
The proposed video streaming architecture is based on a combination of network coding and multi connectivity with \gls{lte}, which are managed by a middle layer between the application and the transport layers.
Multi connectivity avoids service unavailability during mmWave outages, making it possible to support uninterrupted video streaming, while network coding introduces additional robustness against packet loss and simplifies the management of multiple data paths.

The proposed solution was evaluated using a novel simulation framework which joins the ns-3 mmWave module with a realistic application layer based on real H.264 video traces.
This framework allowed us to investigate the application layer performance using typical video metrics such as the \gls{nalu} loss ratio and the \gls{psnr}.
The results confirm the benefit introduced by multi connectivity, and show that network coding can help reduce the \gls{nalu} loss and increase the \gls{psnr}, especially when the mmWave-only solution is used. 

As part of our future work, we will study how it is possible to achieve additional performance gain with network coding, by investigating additional parameter configurations in order to decrease the \gls{nalu} loss without increasing the latency.
Moreover, we will extend the simulation framework to decode also the spatial enhancement layer and better characterize the \gls{psnr} of the received video.

\bibliographystyle{IEEEtran}
\bibliography{bibl.bib}

\begin{thebibliography}{10}
\providecommand{\url}[1]{#1}
\csname url@samestyle\endcsname
\providecommand{\newblock}{\relax}
\providecommand{\bibinfo}[2]{#2}
\providecommand{\BIBentrySTDinterwordspacing}{\spaceskip=0pt\relax}
\providecommand{\BIBentryALTinterwordstretchfactor}{4}
\providecommand{\BIBentryALTinterwordspacing}{\spaceskip=\fontdimen2\font plus
\BIBentryALTinterwordstretchfactor\fontdimen3\font minus
  \fontdimen4\font\relax}
\providecommand{\BIBforeignlanguage}[2]{{%
\expandafter\ifx\csname l@#1\endcsname\relax
\typeout{** WARNING: IEEEtran.bst: No hyphenation pattern has been}%
\typeout{** loaded for the language `#1'. Using the pattern for}%
\typeout{** the default language instead.}%
\else
\language=\csname l@#1\endcsname
\fi
#2}}
\providecommand{\BIBdecl}{\relax}
\BIBdecl

\bibitem{cisco}
Cisco, ``Cisco visual networking index: Forecast and methodology,
  2016–2021,'' \emph{White paper, Cisco public}, 2017.

\bibitem{zeqi2017furion}
Z.~Lai, Y.~C. Hu, Y.~Cui, L.~Sun, and N.~Dai, ``Furion: Engineering
  high-quality immersive virtual reality on today's mobile devices,'' in
  \emph{Proceedings of the 23rd Annual International Conference on Mobile
  Computing and Networking}, ser. MobiCom '17, 2017, pp. 409--421.

\bibitem{rangan2014potential}
S.~Rangan, T.~S. Rappaport, and E.~Erkip, ``Millimeter-wave cellular wireless
  networks: Potentials and challenges,'' \emph{Proceedings of the IEEE}, vol.
  102, no.~3, pp. 366--385, March 2014.

\bibitem{lu2012modeling}
J.~S. Lu, D.~Steinbach, P.~Cabrol, and P.~Pietraski, ``{Modeling human blockers
  in millimeter wave radio links},'' \emph{ZTE Communications}, vol.~10, no.~4,
  pp. 23--28, 2012.

\bibitem{mezzavilla2017end}
\BIBentryALTinterwordspacing
M.~Mezzavilla, M.~Zhang, M.~Polese, R.~Ford, S.~Dutta, S.~Rangan, and M.~Zorzi,
  ``{End-to-End Simulation of 5G mmWave Networks},'' \emph{Submitted to IEEE
  Communication Surveys \& Tutorials}, 2017. [Online]. Available:
  \url{https://arxiv.org/abs/1705.02882}
\BIBentrySTDinterwordspacing

\bibitem{pedersen2011kodo}
M.~V. Pedersen, J.~Heide, and F.~H. Fitzek, ``Kodo: An open and research
  oriented network coding library,'' in \emph{International Conference on
  Research in Networking}.\hskip 1em plus 0.5em minus 0.4em\relax Springer,
  2011, pp. 145--152.

\bibitem{measy}
\BIBentryALTinterwordspacing
Measy, ``{60 GHz video transmitter},'' 2017. [Online]. Available:
  \url{http://www.measy.com.cn/product/showproduct143_en.htm}
\BIBentrySTDinterwordspacing

\bibitem{nitsche2014ieee}
T.~Nitsche, C.~Cordeiro, A.~B. Flores, E.~W. Knightly, E.~Perahia, and J.~C.
  Widmer, ``{IEEE 802.11 ad: directional 60 GHz communication for
  multi-Gigabit-per-second Wi-Fi},'' \emph{IEEE Communications Magazine},
  vol.~52, no.~12, pp. 132--141, December 2014.

\bibitem{singh2008video}
H.~Singh, J.~Oh, C.~Kweon, X.~Qin, H.~R. Shao, and C.~Ngo, ``{A 60 GHz wireless
  network for enabling uncompressed video communication},'' \emph{IEEE
  Communications Magazine}, vol.~46, no.~12, pp. 71--78, December 2008.

\bibitem{abari17enabling}
O.~Abari, D.~Bharadia, A.~Duffield, and D.~Katabi, ``Enabling high-quality
  untethered virtual reality,'' in \emph{14th USENIX Symposium on Networked
  Systems Design and Implementation (NSDI 17)}.\hskip 1em plus 0.5em minus
  0.4em\relax USENIX Association, pp. 531--544.

\bibitem{vukobratovic2014random}
D.~Vukobratović, C.~Khirallah, V.~Stanković, and J.~S. Thompson, ``{Random
  Network Coding for Multimedia Delivery Services in LTE/LTE-Advanced},''
  \emph{IEEE Transactions on Multimedia}, vol.~16, no.~1, pp. 277--282, January
  2014.

\bibitem{jin2008is}
J.~Jin, B.~Li, and T.~Kong, ``{Is Random Network Coding Helpful in WiMAX?}'' in
  \emph{IEEE INFOCOM 2008 - The 27th Conference on Computer Communications},
  April 2008.

\bibitem{tassi2015resource}
A.~Tassi, C.~Khirallah, D.~Vukobratović, F.~Chiti, J.~S. Thompson, and
  R.~Fantacci, ``{Resource Allocation Strategies for Network-Coded Video
  Broadcasting Services Over LTE-Advanced},'' \emph{IEEE Transactions on
  Vehicular Technology}, vol.~64, no.~5, pp. 2186--2192, May 2015.

\bibitem{song2012performance}
W.~Song and W.~Zhuang, ``Performance analysis of probabilistic multipath
  transmission of video streaming traffic over multi-radio wireless devices,''
  \emph{IEEE Transactions on Wireless Communications}, vol.~11, no.~4, pp.
  1554--1564, April 2012.

\bibitem{mmMAGIC}
H.~Halbauer, P.~Rugelandand, R.~Tanoand, M.~Tercero, A.~Vijay, Y.~Li,
  M.~Filippouand, H.~Miao, J.~Widmerand, C.~Fiandrino, I.~Siaudand, A.-M.
  Ulmer-Moll, M.~Shariat, J.~Lorcaand, and Y.~Zou, ``{Evaluations of the
  concepts for the 5G architecture and integration},'' mmMAGIC Deliverable
  D3.2, June 2017.

\bibitem{stockhammer2011dash}
T.~Stockhammer, ``{Dynamic Adaptive Streaming over HTTP --: Standards and
  Design Principles},'' in \emph{Proceedings of the Second Annual ACM
  Conference on Multimedia Systems}, ser. MMSys '11, 2011, pp. 133--144.

\bibitem{RFC7826}
H.~Schulzrinne, A.~Rao, R.~Lanphier, M.~Westerlund, and M.~Stiemerling,
  ``{Real-Time Streaming Protocol Version 2.0},'' Internet Requests for
  Comments, RFC 7826, December 2016.

\bibitem{zhang2016transport}
M.~Zhang, M.~Mezzavilla, R.~Ford, S.~Rangan, S.~Panwar, E.~Mellios, D.~Kong,
  A.~Nix, and M.~Zorzi, ``{Transport layer performance in 5G mmWave
  cellular},'' in \emph{IEEE Conference on Computer Communications
  Millimeter-wave Networking Workshop (INFOCOM WKSHPS)}, 2016, pp. 730--735.

\bibitem{polese2017tcp}
M.~Polese, R.~Jana, and M.~Zorzi, ``{TCP and MP-TCP in 5G mmWave Networks},''
  \emph{IEEE Internet Computing}, vol.~21, no.~5, pp. 12--19, September 2017.

\bibitem{raiciu2011improving}
C.~Raiciu, S.~Barre, C.~Pluntke, A.~Greenhalgh, D.~Wischik, and M.~Handley,
  ``{Improving Datacenter Performance and Robustness with Multipath TCP},'' in
  \emph{Proceedings of the ACM SIGCOMM 2011 Conference}, ser. SIGCOMM '11,
  2011, pp. 266--277.

\bibitem{polese2017improved}
M.~Polese, M.~Giordani, M.~Mezzavilla, S.~Rangan, and M.~Zorzi, ``{Improved
  Handover Through Dual Connectivity in 5G mmWave Mobile Networks},''
  \emph{IEEE Journal on Selected Areas in Communications}, vol.~35, no.~9, pp.
  2069--2084, September 2017.

\bibitem{dasilva}
I.~D. Silva, G.~Mildh, J.~Rune, P.~Wallentin, J.~Vikberg, P.~Schliwa-Bertling,
  and R.~Fan, ``{Tight Integration of New {5G} Air Interface and {LTE} to
  Fulfill {5G} Requirements},'' in \emph{IEEE 81st Vehicular Technology
  Conference (VTC Spring)}, May 2015.

\bibitem{rfc6824}
A.~Ford, C.~Raiciu, M.~Handley, and O.~Bonaventure, ``{TCP Extensions for
  Multipath Operation with Multiple Addresses},'' RFC 6824, 2013.

\bibitem{ahlswede2000network}
R.~Ahlswede, N.~Cai, S.~Y.~R. Li, and R.~W. Yeung, ``Network information
  flow,'' \emph{IEEE Transactions on Information Theory}, vol.~46, no.~4, pp.
  1204--1216, July 2000.

\bibitem{ho2006random}
T.~Ho, M.~Medard, R.~Koetter, D.~R. Karger, M.~Effros, J.~Shi, and B.~Leong,
  ``A random linear network coding approach to multicast,'' \emph{IEEE
  Transactions on Information Theory}, vol.~52, no.~10, pp. 4413--4430, October
  2006.

\bibitem{NC-areview}
E.~Magli, M.~Wang, P.~Frossard, and A.~Markopoulou, ``Network coding meets
  multimedia: A review,'' \emph{IEEE Transactions on Multimedia}, vol.~15,
  no.~5, pp. 1195--1212, August 2013.

\bibitem{heide2011code}
J.~Heide, M.~V. Pedersen, F.~H. Fitzek, and M.~M{\'e}dard, ``{On code
  parameters and coding vector representation for practical RLNC},'' in
  \emph{IEEE International Conference on Communications (ICC)}, 2011.

\bibitem{avc}
T.~Wiegand, G.~J. Sullivan, G.~Bjontegaard, and A.~Luthra, ``{Overview of the
  H.264/AVC video coding standard},'' \emph{IEEE Transactions on Circuits and
  Systems for Video Technology}, vol.~13, no.~7, pp. 560--576, July 2003.

\bibitem{SVC}
H.~Schwarz, D.~Marpe, and T.~Wiegand, ``{Overview of the Scalable Video Coding
  Extension of the H.264/AVC Standard},'' \emph{IEEE Transactions on Circuits
  and Systems for Video Technology}, vol.~17, no.~9, pp. 1103--1120, September
  2007.

\bibitem{UEP-RLC}
D.~Vukobratovic and V.~Stankovic, ``Unequal error protection random linear
  coding strategies for erasure channels,'' \emph{IEEE Transactions on
  Communications}, vol.~60, no.~5, pp. 1243--1252, May 2012.

\bibitem{wns3_16}
R.~Ford, M.~Zhang, S.~Dutta, M.~Mezzavilla, S.~Rangan, and M.~Zorzi, ``{A
  framework for end-to-end evaluation of 5G mmWave cellular networks in
  ns-3},'' in \emph{Proceedings of the Workshop on ns-3}.\hskip 1em plus 0.5em
  minus 0.4em\relax ACM, 2016, pp. 85--92.

\bibitem{zhang2017ns}
M.~Zhang, M.~Polese, M.~Mezzavilla, S.~Rangan, and M.~Zorzi, ``{ns-3
  Implementation of the 3GPP MIMO Channel Model for Frequency Spectrum above 6
  GHz},'' in \emph{Proceedings of the Workshop on ns-3}.\hskip 1em plus 0.5em
  minus 0.4em\relax ACM, 2017, pp. 71--78.

\bibitem{JSVM}
\BIBentryALTinterwordspacing
R.~Julien, H.~Schwarz, and M.~Wien, ``{Joint Scalable Video Model 9.19.15 (JSVM
  9.19.15)},'' 2011. [Online]. Available:
  \url{https://github.com/floriandejonckheere/jsvm}
\BIBentrySTDinterwordspacing

\bibitem{SVEF}
A.~Detti, G.~Bianchi, C.~Pisa, F.~S. Proto, P.~Loreti, W.~Kellerer,
  S.~Thakolsri, and J.~Widmer, ``{SVEF: an open-source experimental evaluation
  framework for H.264 scalable video streaming},'' in \emph{2009 IEEE Symposium
  on Computers and Communications}, July 2009, pp. 36--41.

\bibitem{ffmpeg}
\BIBentryALTinterwordspacing
``{FFmpeg},'' 2017. [Online]. Available: \url{https://ffmpeg.org}
\BIBentrySTDinterwordspacing

\end{thebibliography}

\end{document}